\documentclass[10pt,conference,epsfig]{IEEEtran}
\usepackage{amsmath,amsfonts}
\usepackage{algorithmic}
\usepackage{algorithm}
\usepackage{array}
\usepackage{textcomp}
\usepackage{stfloats}
\usepackage{url}
\usepackage{verbatim}
\usepackage{graphicx}
\usepackage{xcolor}
\usepackage{cite}
\usepackage{xcolor}
\usepackage[caption=false,font=normalsize,labelfont=sf,textfont=sf]{subfig}
\usepackage{geometry}
\usepackage{siunitx}
\usepackage{xcolor}

\usepackage{hyperref}

\begin{document}

\title{Topology-Aware Handover Management for Service Continuity in LEO Satellite Networks}

% \author{Daniel ,~\IEEEmembership{Staff,~IEEE,}}

\author{
    \IEEEauthorblockN{
        Jui-Che Han, Nasir Khan, Siva Satya Sri Ganesh Seeram, Cicek Cavdar
    }
    \IEEEauthorblockA{
        Department of Communication Systems,
        KTH Royal Institute of Technology, Sweden\\
        Email: \{jchan2, nkha, sssgse, cavdar\}@kth.se
    }
}
        % <-this % stops a space
%\thanks{This paper was produced by the IEEE Publication Technology Group. They are in Piscataway, NJ.}% <-this % stops a space
%\thanks{Manuscript received April 19, 2021; revised August 16, 2021.}}



\maketitle

\begin{abstract}
Low Earth orbit (LEO) satellites are expected to play a key role in sixth-generation (6G) non-terrestrial networks. The high mobility of LEO satellites necessitates frequent handovers (HOs), posing challenges to service continuity and availability for user equipment. Conventional methods based on handover margin (HOM) and time-to-trigger (TTT) have limited adaptability to time-varying channel conditions and do not explicitly account for the impact of feeder-link and ground station (GS) connectivity on HO interruption time. To address these limitations, this paper develops a topology-aware HO framework based on proximal policy optimization (PPO). The framework uses reference signal received power (RSRP), satellite--GS associations, remaining visibility time, and elevation angles to determine when to initiate an HO and which satellite to select. Simulation results demonstrate that the proposed policy achieves a service availability of 99.32\% at a minimum elevation angle of 40 degrees, exceeding that of optimized HOM--TTT (O-HT) and least-delay (LD) satellite selection by 1.67 and 1.41 percentage points, respectively. The policy maintains service availability above 97.52\% as the minimum elevation angle increases to 55 degrees, demonstrating its ability to sustain high service availability under stricter elevation constraints.
% Low Earth orbit (LEO) satellites are expected to play a key role in sixth-generation (6G) non-terrestrial networks. 
% The high mobility of LEO satellites requires frequent handovers (HOs), making it difficult to maintain service continuity and availability for user equipment.
% Conventional methods leveraging  handover margin (HOM) and time-to-trigger (TTT)  struggle to adapt to time-varying channels and do not account for the service continuity, feeder links and ground station connectivity. 
% To address these challenges, this paper develops a topology-aware HO framework based on proximal policy optimization (PPO) method.
% The framework leverages reference signal received power (RSRP), satellite–ground station (GS) associations, remaining visibility time and elevation angles  to determine when to HO and which satellite to select. Simulation results demonstrate that the proposed policy achieves service availability of 99.32\% under a minimum elevation angle constraint of $40^\circ$, outperforming optimized HOM-TTT (O-HT) by 1.67 percentage points and least-delay satellite selection (LD) by 1.41 percentage points, respectively. The proposed policy further maintains service availability above 97.52\% as the minimum elevation angle increases to $55^\circ$, demonstrating  stable service availability under different elevation constraints. 
\end{abstract}


\section{Introduction}
Satellite communications are a key enabler of ubiquitous connectivity in sixth-generation (6G) networks, extending coverage to remote regions beyond terrestrial coverage \cite{Giordani2021NTN}. Low Earth orbit (LEO) satellites offer lower propagation delays than geostationary systems \cite{Kodheli2021Satellite}, but their rapid motion limits visibility periods and requires frequent inter-satellite handovers (HOs), challenging service continuity \cite{10946034}. Although quasi-Earth-fixed beams i.e., steerable
beams targeting fixed location, reduce HO frequency compared with Earth-moving beams (non-steerable beam) \cite{3GPP38300}, transitions between satellites remain necessary as serving satellites move out of visibility. 
%These transitions introduce signaling overhead and service interruptions that can reduce throughput and degrade connection quality. 
Effective HO management is therefore essential to sustain effective service time, i.e.,  the active service time excluding HO delays, radio link failures (RLFs), and coverage gaps (CGs). 

Recent studies have investigated  conditional HO (CHO) optimization of handover margin (HOM) and time-to-trigger (TTT) threshold to balance HO  frequency against link continuity \cite{10946034}, \cite{DT}. Increasing these thresholds can suppress unnecessary HOs, but may also delay necessary transitions and increase RLFs. 
Moreover, Fixed thresholds struggle to adapt to rapidly changing satellite and channel conditions.  The highly dynamic nature of LEO satellite networks calls for adaptive HO decisions that respond to evolving link and resource conditions in real time \cite{lee2025multi}. 

% In large-scale LEO constellations, evaluating HO conditions across numerous candidate satellites can also incur substantial computational overhead, limiting their suitability for highly dynamic satellite networks.

Reinforcement learning (RL) addresses this need by learning policies that map observed network states to HO actions while accounting for the long-term effects of these decisions \cite{11419899, 10551688}. Existing studies have applied this approach to satellite selection using link and resource states \cite{11112684}, joint HO and power allocation \cite{11435637}, and hierarchical HO timing and target selection \cite{11419899}.
Centralized and distributed RL schemes have also modeled line-of-sight (LoS), shadowed, and blocked links and HO interruption \cite{10551688}. 
Nevertheless, these RL schemes do not explicitly distinguish candidate HOs by topology-dependent signaling delay determined by satellite–GS connectivity. Such differences can affect the duration of service interruption associated with each HO. Therefore, minimizing HO frequency alone does not necessarily minimize cumulative HO interruption time.

To effectively address the problems of joint optimization of HO counts and end-to-end HO delays under different HO topologies, we propose a novel topology-aware satellite HO strategy based on proximal policy optimization (PPO). Our main contributions are summarized as follow:
\begin{itemize}
\item We formulate a joint optimization problem for inter-satellite HO frequency and CHO delay in gNB-onboard LEO networks, incorporating topology-dependent signaling delays under received-power, remaining-visibility-time, and satellite–GS connectivity constraints.

\item We develop a PPO-based satellite-selection policy using received power, elevation angle, remaining visibility time, GS association, and association history. A surrogate reward balances HO overhead and link continuity by penalizing topology-dependent switching costs, short inter-HO intervals, and service interruptions.

\item We assess the effects of end-to-end CHO delay, HO count, and RLF duration on service availability under urban propagation conditions. Compared with optimized HOM-TTT and least-delay baselines, the proposed topology-aware policy achieves over 99.3\% service availability and reduces RLF outage slots by at least 85\%.

\end{itemize}

% The rest of the paper is organized as follows. Section \ref{sec:sys_mod} presents the system model.
% Section \ref{sec:problem_formulation} describes the optimization problem. Section \ref{sec:ho_methods} describes RL-based HO scheme.
% Section \ref{sec:result} evaluates the performance of the proposed.
% Conclusions are presented in Section \ref{sec:con}.

\section{System model}
\label{sec:sys_mod}
\begin{figure}[!t]
\centering
\includegraphics[width=0.5\textwidth]{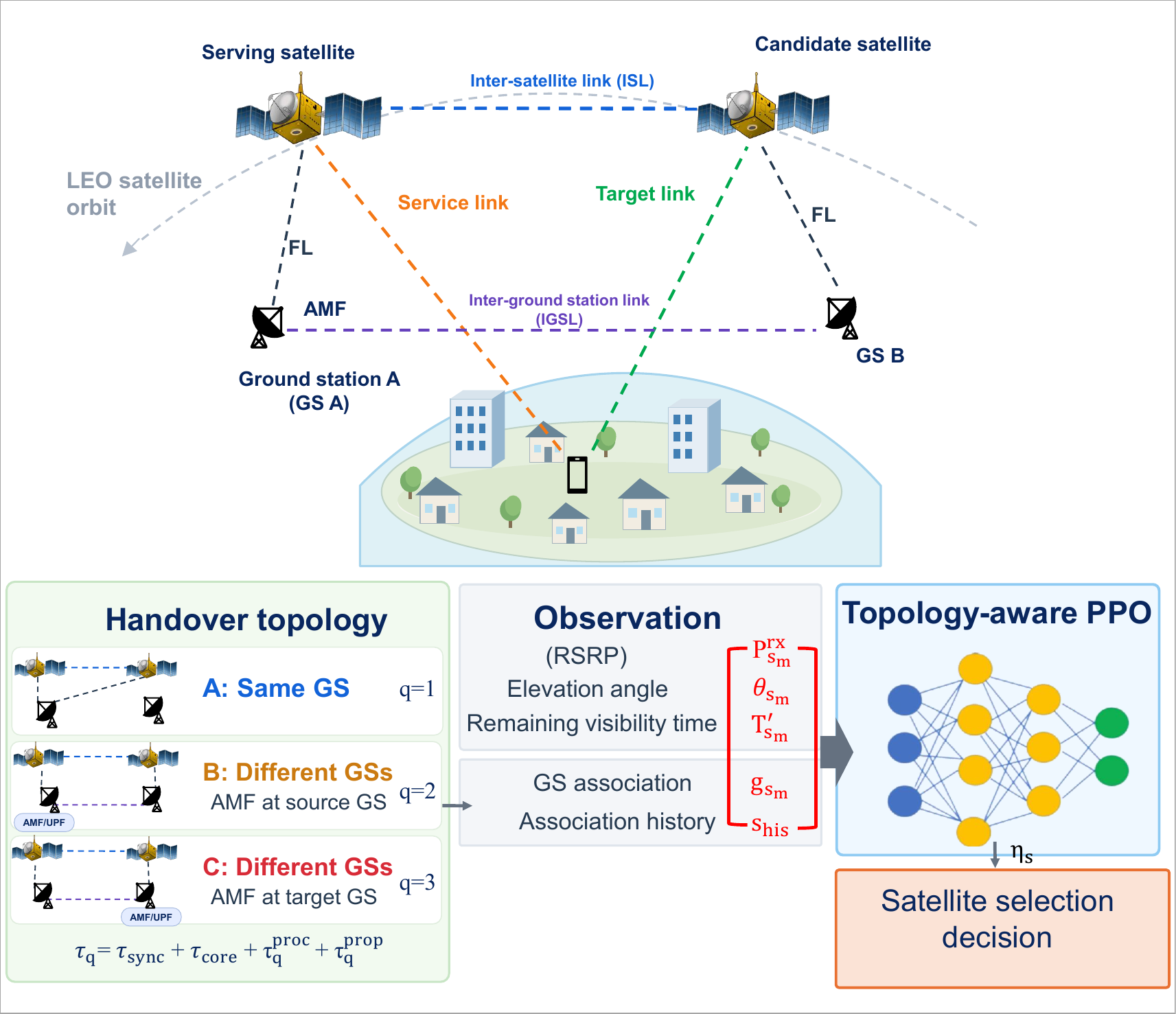}
\caption{System model and topology-aware handover framework for a LEO satellite network.}
\label{model_fig}
\end{figure}

As shown in Fig.\ref{model_fig}, the network comprises a ground based user equipment (UE), a LEO satellite constellation $\mathcal{S}=\{1,\ldots,S\}$, and ground stations (GSs) $\mathcal{G}=\{1,\ldots,G\}$, observed over $T$ slots of duration $\Delta t$
Each satellite hosts gNB functionality following the gNB-onboard regenerative architecture \cite{3gpp.38.821}. 
The UE–satellite and satellite–GS connections use service links (SLs) and feeder links (FLs), respectively, while inter-satellite links (ISLs) and inter-GS links (IGSLs) carry HO signaling.
Each satellite uses a directional antenna with an equivalent circular aperture and quasi-Earth-fixed beams under ideal beam steering. We consider only inter-satellite HOs, with the UE remaining on the same beam during each serving interval.
Let $\theta_s(t)$ and $\beta_{s,j}(t)$ denote the elevation angle of satellite $s$ as seen from the UE and GS $j$ at time $t$, respectively, measured from the local horizontal. Under unobstructed LoS conditions, the elevation angle thresholds $\theta_{\min}$ and $\beta_{\min}$ determine geometric visibility at the UE and GSs. For satellite $s$, the set of visible GSs at time slot $t$ is
$    \mathcal{G}^{\mathrm{vis}}_s(t)
    =
    \left\{
        j\in\mathcal{G}
        \,\middle|\,
        \beta_{s,j}(t)\geq\beta_{\min}
    \right\}.$
Its direct FL association is represented by  $g_s(t)\in\mathcal{G}^{\mathrm{vis}}_s(t).$ Each satellite satisfying $\theta_{min}$ is assigned to the highest-elevation visible GS with spare capacity; otherwise, lower-elevation visible GSs are considered in descending order. Each GS supports at most two FLs. Satellites satisfying both UE visibility and direct FL availability form the eligible set
\begin{equation}
    \mathcal{S}_{\mathrm{av}}(t)
    =
    \left\{
        s\in\mathcal{S}
        \,\middle|\,
        \theta_s(t)\geq\theta_{\min},
        \; g_s(t) \in \mathcal{G}^{\mathrm{vis}}_s(t)
    \right\}.
    \label{eq:available_satellites}
\end{equation}
Let $\eta(t)\in\mathcal{S}_{\mathrm{av}}(t)\cup{\{0\}}$ denote the UE–satellite association in time slot $t$, where $\eta(t)=s$ indicates the UE is connected with satellite $s$, and $\eta(t)=0$ denotes no association or an RLF. Thus, a HO is counted only when the serving satellite changes from one satellite to another, excluding transitions to or from the unassociated state.

\subsection{Propagation and Received-Power Models}
Following the large-scale propagation model in \cite{3gpp.38.811}, we consider an outdoor UE under clear-sky and LoS conditions, neglecting clutter loss, building penetration loss, atmospheric attenuation, and scintillation to simplify our analysis. Shadow fading is retained to capture propagation variability across environments. The total path loss, in dB, is approximated as
\begin{equation}
    \mathrm{PL}_s(t)
    \approx PL_{b,s}(t)
    = PL_{\mathrm{FS},s}(t) + SF_s(t),
    \label{eq:basic_path_loss}
\end{equation}
where $PL_{b,s}(t)$ denotes the basic path loss, and
$PL_{\mathrm{FS},s}(t)$ and $SF_s(t)$ denote the free-space
path loss (FSPL) and shadow fading, respectively, modeled
according to \cite{3gpp.38.811}.

Following the link-budget model in \cite{3gpp.38.821}, the
downlink received power from satellite $s$, in dBW, is
\begin{equation}
    P^{\mathrm{rx}}_s(t)
    =
    \operatorname{EIRP}_s + G_r - \mathrm{PL}_s(t),
    \label{eq:received_power}
\end{equation}
where $\operatorname{EIRP}_s = P_s^{tx} + G_{t,s}$ is the
equivalent isotropically radiated power in dBW, $P_s^{tx}$
is the transmit power in dBW, and $G_{t,s}$ and $G_r$ are
the transmit and receive-antenna gains in dBi, respectively.
The antenna gains are assumed constant over the considered
links.

\subsection{Topologies and Handover Delay Model }

We adapt the CHO delay model in \cite{seeram2025handover} to inter-satellite HOs with direct FLs. Each HO transfers the UE connection from a source satellite to a target satellite. Depending on their GS associations and the location of the UE's core-network functions, we consider three mobility scenarios:

\subsubsection{Topology A}
The source and target satellites have direct FLs to the same GS. The HO changes the serving satellite while retaining the GS connection.

\subsubsection{Topology B}
The source and target satellites have direct FLs to different GSs. The UE's access and mobility management function (AMF) and user plane function (UPF) are hosted at the source GS. The HO therefore changes both the serving satellite and the GS used to reach these core-network functions, which remain at the source GS.

\subsubsection{Topology C}
The source and target satellites have direct FLs to different GSs, with the UE's AMF and UPF hosted at the target GS. The HO changes the serving satellite and connects the UE through the target satellite's direct FL to the GS hosting these core-network functions.
    
% These topologies use a common AMF/UPF anchor, whereas the inter-AMF mobility is not modeled. 

To characterize HO frequency and delay in time slot $t$, we define the HO triggering by
 \begin{equation}
 h(t)= \mathbb I\{\eta(t-1)\ne0,\eta(t)\ne0,\ \eta(t)\ne \eta(t-1)\}.
     \label{HO:happen}
\end{equation}
Then, the total number of HOs over $T$ time slots is given by
\begin{equation}
    N_{\rm HO}=\sum_{t=2}^{T}h_t,
    \label{HO:count}
\end{equation}
Under the gNB-onboard architecture, we restrict inter-satellite HOs to the modeled topologies
$\mathcal{Q}=\{\mathrm{A},\mathrm{B},\mathrm{C}\}$, 
% Let $z_q(t)\in\{0,1\}$ indicate whether the HO at slot $t$ corresponds to topology $q$, such that
% \begin{equation}
%     \sum_{q\in\mathcal{Q}}z_q(t)=h_t.
% \end{equation}
and each HO could only be assign to one topology. % and sets all topology indicators to zero otherwise.
The HO delay associated with topology $q$ at time slot $t$ is expressed as \cite{seeram2025handover}
\begin{equation}
    \tau_{\mathrm{CHO}}(t)
    =
    \tau_{\mathrm{sync}}
    +\tau_{\mathrm{core}}
    +\tau_q^{\mathrm{proc}}
    +\tau_q^{\mathrm{prop}}(t),
\end{equation}
where the four terms represent synchronization, additional core-network, processing, and propagation delays, respectively. When no HO occurs, i.e., $h(t)=0$, we set $\tau_{\mathrm{CHO}}(t)=0$. Assuming a common per-message processing delay $\tau^{\mathrm{ppm}}$, the processing delay and propagation delay is
\begin{equation}
    \tau_q^{\mathrm{proc}}
    =
    \tau^{\mathrm{ppm}}
    \sum_{\ell\in\mathcal{L}}N_q^\ell,  \tau_q^{\mathrm{prop}}(t)
    =
    \sum_{\ell\in\mathcal{L}}
    N_q^\ell\,\bar{\tau}_q^\ell(t),
    \label{eq:processing_delay}
\end{equation}
repectively, where $\mathcal{L}=\{\mathrm{FL},\mathrm{SL},\mathrm{ISL},\mathrm{IGSL}\}$ denotes the set of link types, $N_q^\ell$ is the number of signaling-message transmissions over type-$\ell$ links in topology $q$ \cite{seeram2025handover}, and $\bar{\tau}_q^\ell(t)$ denotes the mean one-way propagation delay across the $N_q^\ell$ type-$\ell$ transmissions, accounting for differences in source and target link lengths. Terms with $N_q^\ell=0$ contribute zero. Link geometry is assumed constant during each HO procedure. 
% The realized HO delay at slot $t$ is
% \begin{equation}
%     \tau_{\mathrm{CHO}}(t)
%     =
%     \sum_{q\in\mathcal{Q}}z_q(t)\tau_q(t),
% \end{equation}
% which equals zero when no HO occurs. 
The cumulative HO delay over the observation horizon is therefore
\begin{equation}
    D_{\mathrm{HO}}
    =
    \sum_{t=2}^{T}\tau_{\mathrm{CHO}}(t).
\end{equation}

\section{Problem Formulation}
\label{sec:problem_formulation}
% For a given O-RAN functional split and Earth-fixed beam deployment, we optimize satellite association to balance received power, HO frequency, and HO delay. At each time slot, the
% association decision determines whether the UE retains its
% current satellite or selects another satellite.
To jointly  minimize
inter-satellite HO frequency and CHO delay, the joint optimization problem is formulated as
\begin{subequations}
\label{eq:optimization_problem}
\begin{align}
\underset{\{\eta(t)\}^T_{t=1}}{\operatorname{min}}
\quad
{}&
\omega_1
\frac{N_{HO}}
     {T\Delta t} +
\omega_2
 \frac{D_{\mathrm{HO}}}{T \Delta t},
\label{eq:objective}\\
\text{s.t.}\quad
&P^{\mathrm{rx}}_{\eta(t)}(t)
\geq P_{\mathrm{th}},
\quad \forall t\;
: \eta(t) \neq 0
\label{eq:constraint_power}\\
&T^{\prime}_{\eta(t)}(t)
\geq T_{\min}^{\prime},
\quad \forall t\;
: \eta(t) \neq 0
\label{eq:constraint_remaining}\\
& \eta(t)
\in  \mathcal{S}_{\mathrm{av}}(t) \cup \{0\}, \quad \forall t,\;
\label{eq:constraint_available}\\
&\sum_{s\in\mathcal{S}_{\mathrm{av}}(t)\cup \{0\}} \mathbb I\{\eta(t) = s\} = 1, \quad \forall t,
\label{eq:constraint_connection}
\end{align}
\end{subequations}
The nonnegative weights $\omega_1$ and $\omega_2$ balance the penalties on HO frequency and cumulative HO delay, respectively. Constraint \eqref{eq:constraint_power} requires the received power $P_{\eta(t)}^{\mathrm{rx}}(t)$ to meet the threshold $P_{\mathrm{th}}$. Constraint \eqref{eq:constraint_remaining} requires the expected remaining visibility time $T_{\eta(t)}^{\prime}(t)$ to exceed or equal $T_{\min}^{\prime}=\Delta t$, ensuring visibility for at least one additional slot. Constraint \eqref{eq:constraint_available} specifies the set of candidate satellites eligible for selection. constraint \eqref{eq:constraint_connection} enforce the selection of exactly one satellite or remains unassociated in each time slot. Moreover, $\eta(t) \neq 0$ requires the UE to select a satellite from the available set $\mathcal{S}_{\mathrm{av}}(t)$ whenever at least one satellite satisfies all of the received power, remaining visibility time, and visibility requirements.

Problem~\eqref{eq:optimization_problem} involves sequential decision-making for target satellite selection, as current associations affect subsequent HO costs under evolving link conditions and satellite visibility. Conventional  RSRP threshold based HO mechanism incur significant complexity as the scale of network increases. To address these issues, we propose to exploit
DRL to solve the problem as detailed next.  

\section{Reinforcement Learning-Based Handover Scheme}
\label{sec:ho_methods}

\subsection{Problem Transformation into RL Framework}
\label{PPO}
To account for the long-term effects of satellite association, we formulate the HO decision process within an RL framework, where the UE acts as an agent interacting with the satellite network. At each slot $t$, the agent observes $\mathbf{s}(t)$, selects an action $\mathbf{a}(t)$ according to its policy, and receives the reward $\tilde{\mathbf{r}}(t)$ and the next observation $\mathbf{s}(t+1)$. The policy maximizes the expected discounted cumulative reward, thereby accounting for both immediate HO costs and their subsequent consequences.
% \red{The current candidate-satellite observations and a finite association history jointly provide the state representation. We describe this interaction by the MDP model
% $\mathcal{M}=(\mathcal{O},\mathcal{A},\mathcal{P},\mathcal{R},\gamma)$,
% where $\mathcal{O}$ is the observation space used as the state representation, $\mathcal{A}$ is the action space, $\mathcal{P}$ is the observation-transition model conditioned on the current observation and action, $\mathcal{R}$ is the reward function, and $\gamma\in[0,1)$ is the discount factor. This model assumes that the included observations and history sufficiently characterize the decision process.
The observation, action, and reward are defined below.

\subsubsection{Observation Space}
At slot $t$, the observation comprises features of $M$ candidate slots and the UE's association history:
\begin{align}
\mathbf{s}(t)
&=
\left[
\mathbf{s}_{\mathrm{loc},\eta_{1}}(t),\ldots,
\mathbf{s}_{\mathrm{loc},\eta_{M}}(t),
\mathbf{s}_{\mathrm{his}}(t)
\right],
\label{eq:complete_state}\\
\mathbf{s}_{\mathrm{loc},\eta_m}(t)
&=
\left[
P^{\mathrm{rx}}_{\eta_m}(t),
\theta_{\eta_m}(t),
g_{\eta_m}(t),
T^{\prime}_{\eta_m}(t)
\right],
\label{eq:satellite_state}\\
\mathbf{s}_{\mathrm{his}}(t)
&=
\left[
\theta^{\prime}(t-1),
i^{\prime}(t-1),
\bar{T}_{\mathrm{HO}}(t)
\right].
\label{eq:global_state}
\end{align}
identifies the satellite in candidate slot $m$ at time slot $t$, with $\eta_m=0$ denoting padding. Local features comprise RSRP, elevation angle, associated GS, and expected remaining visibility time, respectively. The history records the previous serving satellite's elevation angle $\theta^{\prime}(t-1)$, identifier $i^{\prime}(t-1)$, and elapsed time since the last HO, $\bar{T}_{\mathrm{HO}}(t)$.
The elapsed-time feature is initialized to one and bounded away from zero when evaluating the inverse-time reward penalty. Here, $M$ denotes the maximum number of simultaneously observable satellites, yielding a fixed observation dimension of $4M+3$. Unused slots contain dummy states. During training, candidate slots and their corresponding action mappings are shuffled jointly to reduce slot-position bias.

\subsubsection{Action Space}
At each slot $t$, the action $\mathbf{a}(t) \in \mathcal{A}_t = \{\eta_1, ...,\eta_M\}$ selects one of the $M$ candidate satellite. 
Selecting the previous serving satellite maintains the connection, whereas selecting a different satellite triggers an HO if \eqref{eq:constraint_power}, \eqref{eq:constraint_remaining}, \eqref{eq:constraint_available}, and \eqref{eq:constraint_connection} are satisfied. Selecting a dummy or infeasible candidate incurs a service-interruption penalty.

\subsubsection{Reward Function}
The reward captures topology-dependent HO delay, the spacing between successive HOs, and service interruptions. Let $c$ denote topology occurs between the selected satellite and source satellite. For a feasible HO, the processing-plus-propagation delay is normalized as
\begin{equation}
\bar{\tau}_c(t)
=
\frac{
\tau_c^{\mathrm{proc}}(t)
+
\tau_c^{\mathrm{prop}}(t)
}{
\displaystyle
\max_{c'}
\left(
\tau_{c'}^{\mathrm{proc}}(t)
+
\tau_{c'}^{\mathrm{prop}}(t)
\right)
},
\end{equation}
where the maximization is over feasible HO targets in $\mathcal{S}_{\mathrm{av}}(t)$, excluding the previous serving satellite. The delay terms correspond to the topology between the source satellite and each candidate target. The common synchronization and core-network contribution, $\tau_{\mathrm{sync}}+\tau_{\mathrm{core}}$, is omitted from this target-dependent delay measure. For decisions without an HO, $\bar{\tau}_c(t)$ is set to zero.

To penalize both HO delay and frequent switching \cite{11419899}, the transition reward is defined as
\begin{equation}
\tilde{\mathbf{r}}(t)
=
\begin{cases}
r(t)=
-\alpha_1 h_t
\left(
\bar{\tau}_c(t)
+
\dfrac{1}{\bar{T}_{\mathrm{HO}}(t)}
\right),
& \text{if feasible},\\
-P, & \text{otherwise}.
\end{cases}
\label{eq:reward}
\end{equation}
A feasible selection satisfies the candidate availability conditions in \eqref{eq:constraint_power}, \eqref{eq:constraint_remaining}, \eqref{eq:constraint_available}, and \eqref{eq:constraint_connection}. Invalid selections, RLFs, and CGs incur a penalty of $-P$, with $P$ substantially exceeding the maximum feasible HO penalty, $\max|r(t)|$. RLF penalties are also used in adaptive HO schemes \cite{10949111}.
The inverse-time term uses $\bar{T}_{\mathrm{HO}}(t)$ evaluated before resetting the elapsed-time counter following an HO, thereby penalizing closely spaced HOs. The nonnegative weight $\alpha_1$ controls the HO penalty, and a feasible decision without an HO yields $r(t)=0$. This reward serves as a surrogate for the switching-overhead objective while additionally accounting for service interruptions. Algorithm~\ref{alg:PPO-SHD} summarizes the procedure.

\subsection{Least-Delay Satellite Selection}
\label{sec:md}
We also propose a  least-delay (LD) satellite-selection method. Once the HOM and TTT conditions are met, LD selects the eligible target with the lowest predicted HO delay, replacing the highest-RSRP selection criterion used in \cite{10946034}.

\begin{algorithm}[t]
\caption{PPO for Satellite HO Decisions (PPO-SHD)}
\label{alg:PPO-SHD} \footnotesize
\begin{algorithmic}[1]
\STATE Initialize actor $\pi_\theta$, critic $V_\phi$, buffer
$\mathcal{D}$, and parameters $\gamma,\lambda,\epsilon,H,T$
\FOR{$e=0,\ldots,E-1$}
    \STATE Reset the environment; $t\gets0$
    \WHILE{$t<T$}
        \STATE Construct $\mathbf{s}(t)$ using
        \eqref{eq:complete_state}; add the dummy state
        and shuffle candidate states
        \STATE Sample $\mathbf{a}(t)
        \sim\pi_\theta(\cdot\mid \mathbf{s}(t))$;
        record $\ell_t=\log\pi_\theta(
        \mathbf{a}(t)\mid \mathbf{s}(t))$
        and $v_t=V_\phi(\mathbf{s}(t))$
        \STATE Execute $\mathbf{a}(t)$; observe
        $\tilde{\mathbf{r}}^{(t)}$ from \eqref{eq:reward},
        $\mathbf{s}(t+1)$, and $\mathbf{d}(t)$
        \STATE Store
        $(\mathbf{s}(t),\mathbf{a}(t),\tilde{\mathbf{r}}^{(t)},
        \mathbf{s}(t+1),\mathbf{d}(t),\ell_t,v_t)$ in $\mathcal{D}$
        \STATE $t\gets t+1$
        \IF{$|\mathcal{D}|=H$ or $d^{(t)}=1$ or $t\geq T$}
            \STATE Compute generalized advantage estimation advantages $\hat{A}^{(k)}$
            and returns $\hat{G}^{(k)}$ from $\mathcal{D}$
            \FOR{$j=1,\ldots,K$}
                \STATE Sample minibatch $\mathcal{B}_j$
                from $\mathcal{D}$
                \STATE Compute $r^{(t)}(\theta)$,
                $\bar{\rho}_j^{(t)}$,
                $\mathcal{L}_j^{\mathrm{actor}}$, and
                $\mathcal{L}_j^{\mathrm{critic}}$
                \STATE Update $\theta$ and $\phi$
                using $\mathcal{L}^{clip}$
            \ENDFOR
            \STATE Clear $\mathcal{D}$
        \ENDIF
    \ENDWHILE
\ENDFOR
\end{algorithmic}
\end{algorithm}

\section{Simulation Results}
\label{sec:result}
\subsection{Simulation Setup}
\label{sec}

To simulate satellite--UE communications, we adopt the urban scenario specified in 3GPP TR 38.811 and compute the received power by accounting for FSPL and elevation-dependent shadow fading \cite{3gpp.38.811}. Satellite trajectories determine the time-varying elevation angles and remaining visibility times. The simulation parameters are summarized in Table~\ref{tab:sim}.

For the PPO implementation, the candidate-satellite features and association history defined in Section~\ref{PPO} are concatenated to form a fixed-dimensional observation vector. To promote training stability and efficiency, the candidate-satellite features are normalized prior to neural network processing. Specifically, received powers are converted to linear units and divided by $P_{\mathrm{th}}$, expressed in the same units. Elevation angles and remaining visibility durations are normalized by $90^\circ$ and $T_{\max}^{\prime}$, respectively, where $T_{\max}^{\prime}$ denotes the maximum remaining visibility duration considered in the simulation. The policy network produces a probability distribution over the candidate slots, with the selected slot mapped to the corresponding satellite index.

% \begin{table}[!t]
% \caption{Simulation Parameters}
% \label{tab:sim}
% \centering
% \begin{tabular}{lc}
% \hline
% Parameter & Value \\
% \hline
% LEO constellation \cite{fcc2019spacex} & Starlink 1584 satellites, 72 plane  \\
% Altitude \cite{fcc2019spacex} & 550 km \\
% Inclination \cite{fcc2019spacex} & $53^\circ$ \\
% $\mathcal{G}=\{1,2,3,4\}$ \cite{satellitemap2024starlink} & $\{(38.33458^\circ\mathrm{N},0.4909^\circ\mathrm{E}),$\\
% with 1 and 4 host AMF and UPF & $(50.9871^\circ\mathrm{N},2.1255^\circ\mathrm{E}),$\\
%  & $(45.3207^\circ\mathrm{N},9.1886^\circ\mathrm{E}),$ \\
%  & $(50.3353^\circ\mathrm{N},8.5320^\circ\mathrm{E})\}$ \\
% $\tau^{\mathrm{sync}},\tau^{\mathrm{core}},
% \tau^{\mathrm{ppm}}$ & 20, 50, 1 ms \\
% UE location & $(45.78^\circ\mathrm{N},1.75^\circ\mathrm{E})$ \\
% $T,\Delta t$ & 1200 slots, 1 s \\
% %$a$ & 0.1 m \\
% $P_s^{tx}$ & 23 dBW \\
% $G_{t,s},G_r$ & 30.5 dBi, 20 dBi \\
% $P_{\mathrm{th}}$ & $-109$ dBW \\
% $f_c$ & 20 GHz \\
% $\theta_{\min}$ & $40^\circ$ \\
% Direct FLs per GS & 2 \\
% Shadow-fading environments \cite{3gpp.38.811} & Dense urban, Urban, Rural  \\
% Penalty P & 5 \\
% $\alpha_1$ & 1 \\
% \hline
% \end{tabular}
% \end{table}

\begin{table}[!t]
\footnotesize
\setlength{\tabcolsep}{3pt}
\renewcommand{\arraystretch}{0.95}
\caption{Simulation Parameters}
\label{tab:sim}
\centering
\begin{tabular}{|l|c|}
\hline
Parameter & Value \\
\hline
LEO constellation \cite{fcc2019spacex} & Starlink: 1584 satellites, 72 planes \\
\hline
Altitude \cite{fcc2019spacex} & 550 km \\
\hline
Inclination \cite{fcc2019spacex} & $53^\circ$ \\
\hline
$\mathcal{G}=\{1,2,3,4\}$ \cite{satellitemap2024starlink}
& $\{(38.33458^\circ\mathrm{N},0.4909^\circ\mathrm{E}),$ \\
GSs 1 and 4 host AMF and UPF
& $(50.9871^\circ\mathrm{N},2.1255^\circ\mathrm{E}),$ \\
& $(45.3207^\circ\mathrm{N},9.1886^\circ\mathrm{E}),$ \\
& $(50.3353^\circ\mathrm{N},8.5320^\circ\mathrm{E})\}$ \\
\hline
$\tau^{\mathrm{sync}},\tau^{\mathrm{core}},
\tau^{\mathrm{ppm}}$, $T,\Delta t$ & 20ms, 50ms, 1 ms, 1200 slots, 1 s \\
\hline
UE location & $(45.78^\circ\mathrm{N},1.75^\circ\mathrm{E})$ \\
\hline
%$a$ & 0.1 m \\
$P_{\mathrm{th}}$, $f_c$, $\theta_{\min}$, $\beta_{min}$  &   $-109$ dBW, 20 GHz, $40^\circ$, $10^\circ$ \\
\hline
$\operatorname{EIRP}_s,G_r$ & 53.5dBW, 20 dBi \\
\hline
Direct FLs per GS & 2 \\
\hline
Shadow-fading environments \cite{3gpp.38.811}
& Urban \\
\hline
P, $\alpha_1$, $M$ & 5, 1, 8 \\
\hline
\end{tabular}
\end{table}

\subsection{Performance Metrics}
\label{sec:PM}
We evaluate mobility performance in terms of service continuity and HO stability over an observation period of $T$ time slots. Service interruptions arise from RLFs, CGs, and HO execution. The cumulative outage durations associated with RLFs and CGs are given by $T_{\mathrm{RLF}}=N_{\mathrm{RLF}}\Delta t$ and $T_{\mathrm{CG}}=N_{\mathrm{CG}}\Delta t$, respectively, where $N_{\mathrm{RLF}}$ and $N_{\mathrm{CG}}$ denote the numbers of time slots affected by each outage type. Accounting for these outages and the cumulative HO interruption time $D_{\mathrm{HO}}$, the effective service time is defined as
% \begin{equation}
    $\zeta = T\Delta t - T_{\mathrm{RLF}} - T_{\mathrm{CG}} - D_{\mathrm{HO}}.$
% \end{equation}
Service availability is then expressed as the fraction of the observation period during which service is available:
\begin{equation}
    \mathcal{A} = \frac{\zeta}{T\Delta t}.
\end{equation}

\subsection{Baseline Methods}
We compare the proposed method with two baselines: 
\subsubsection{\textbf{Optimized HOM and TTT (O-HT)}}the baseline adapts from \cite{10946034} to quasi-Earth-fixed beams, using a searched HOM of 15 dB and a TTT of 1 second. 
\subsubsection{\textbf{Least-delay (LD)}}the method described in Section~\ref{sec:md} under the same HOM and TTT settings above.

\subsection{Performance Analysis}

\begin{figure}[t]
    \centering 
    \includegraphics[width=2.6in]{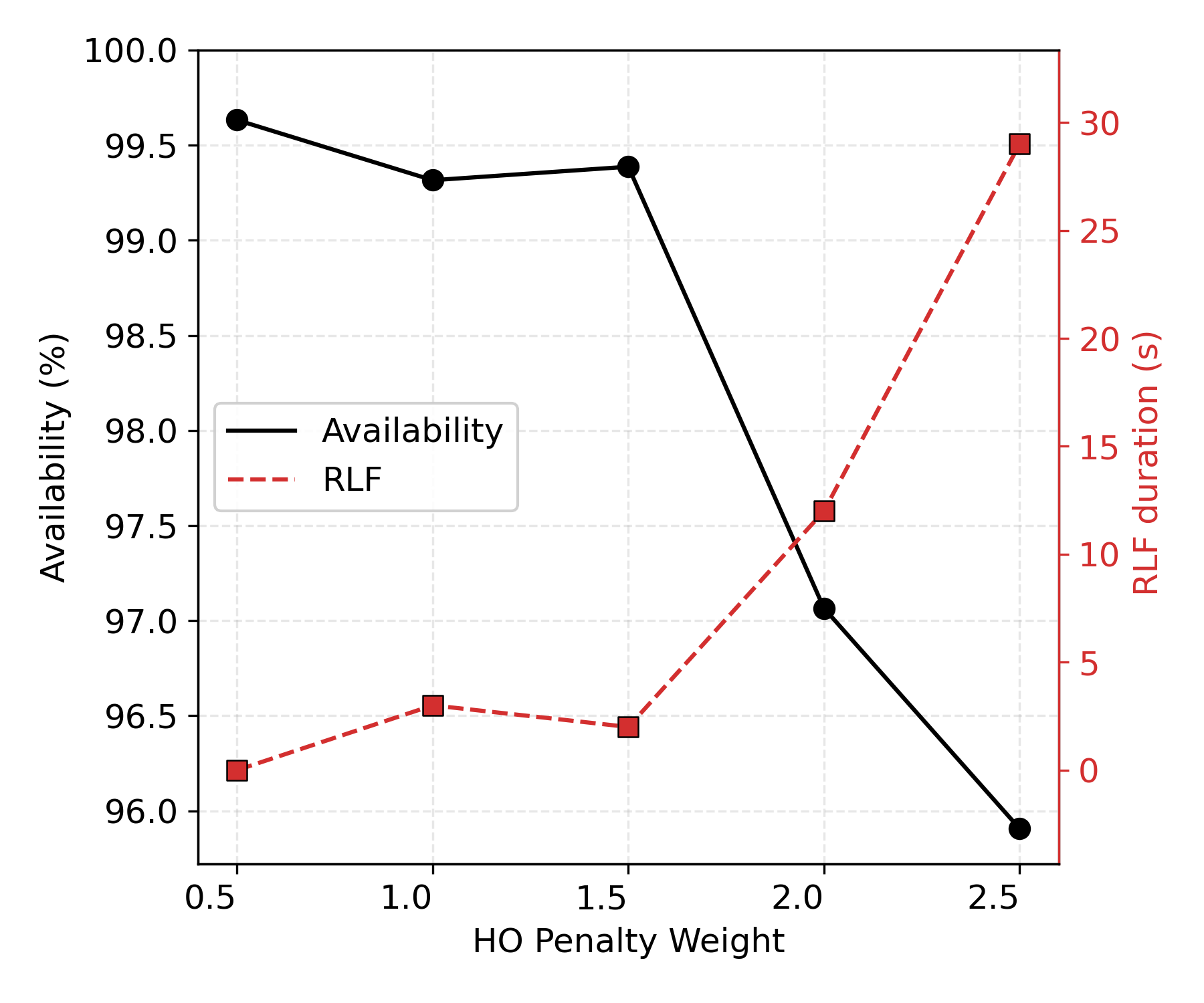} 
    \vspace{-10pt}
    \caption{Effect of the handover penalty weight $\alpha_1$ on service availability and RLF duration.} 
    \label{fig:weight} 
\end{figure}

Fig.~\ref{fig:weight} compares service availability and RLF duration for different HO penalty weights $\alpha_1$. Increasing $\alpha_1$ from 0.5 to 2.5 reduces service availability from 99.63\% to 95.91\% and increases RLF duration from \SI{1}{\second} to \SI{29}{\second}, with more pronounced degradation at higher weights. This trend suggests that stronger HO penalties discourage switching, potentially delaying necessary HOs and increasing the cumulative RLF outage duration. These results highlight the need to balance HO suppression with timely switching to maintain service continuity.
\begin{figure}[t]
    \centering 
    \includegraphics[width=2.6in]{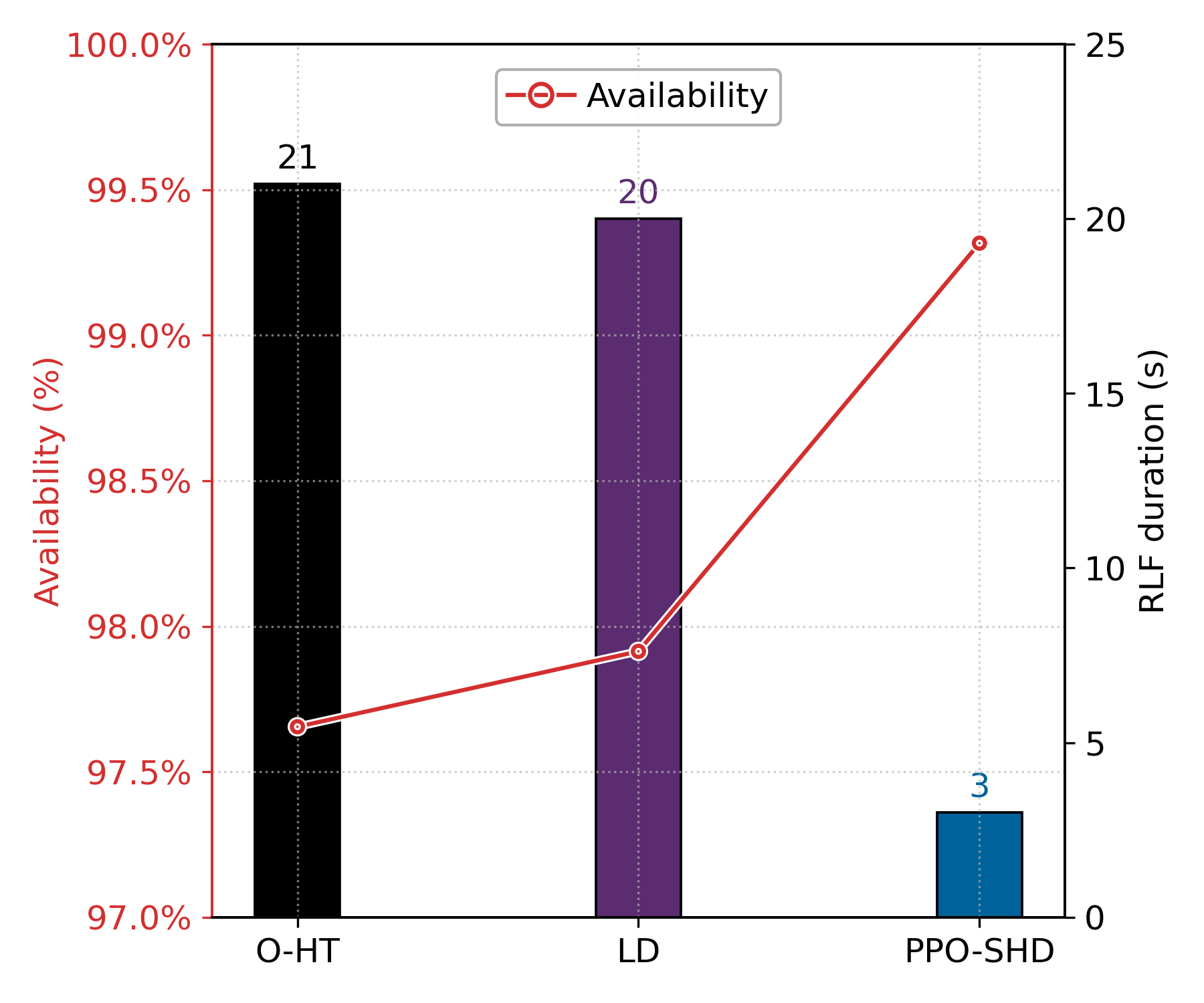} 
       \vspace{-10pt}
    \caption{Comparison of PPO-SHD and baseline performance across availability (red) with RLF duration (bar).} 
    \label{fig:rlf} 
\end{figure}

\begin{figure}[t]
    \centering 
    \includegraphics[width=2.6in]{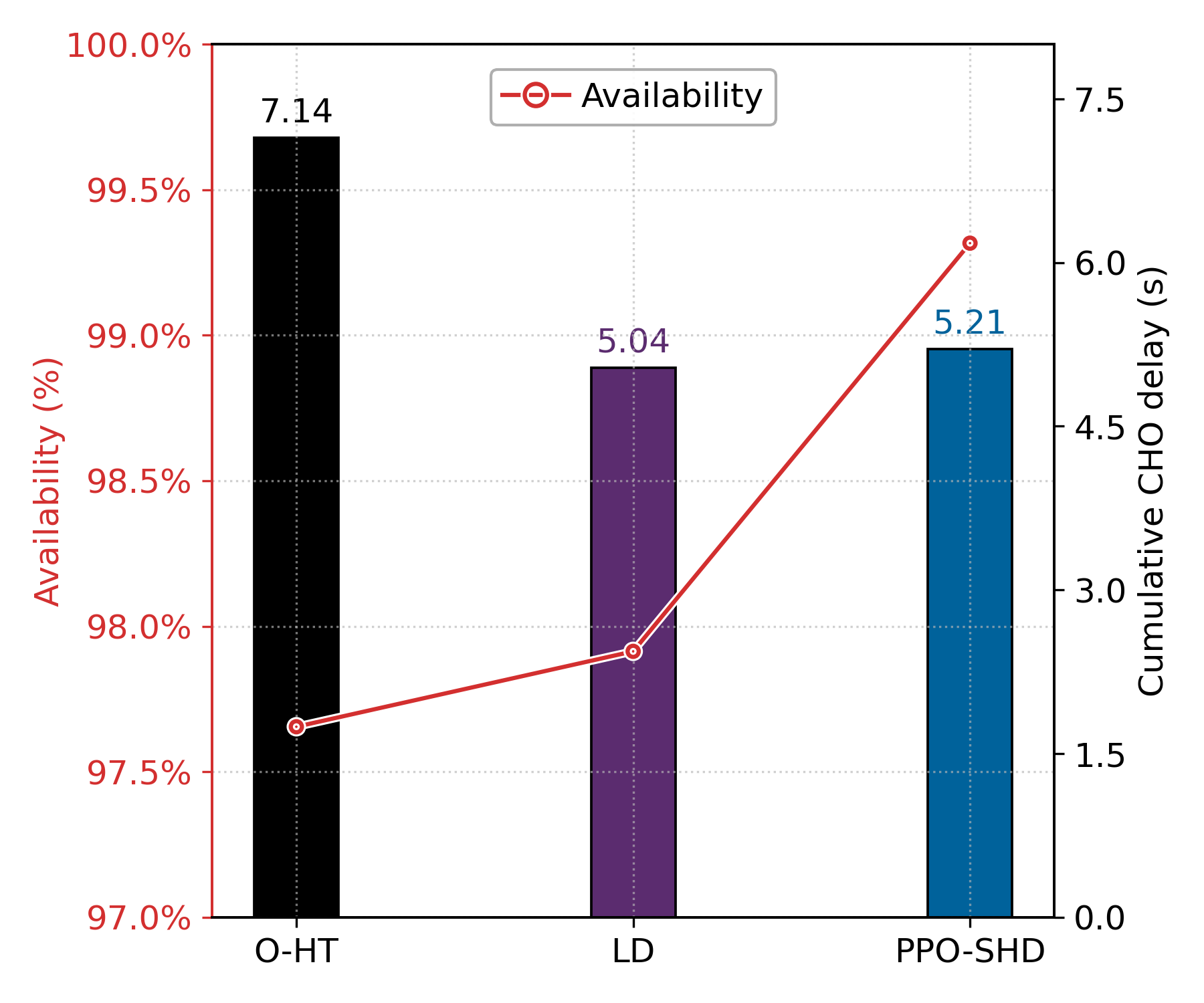} 
        \vspace{-10pt}
    \caption{Comparison of PPO-SHD and baseline performance across availability (red) with cumulative HO time (bar).} 
    \label{fig:cuHO} 
\end{figure}

Fig.~\ref{fig:rlf} illustrates the service availability and RLF duration of our porposed PPO-SHD and the LD and O-HT baselines. The proposed PPO-SHD achieves 99.32\% availability, exceeding O-HT and LD by 1.67 and 1.41 percentage points, respectively. It also reduces RLF duration of 85.7\% and 85.0\%, respectively. These results suggest that PPO-SHD supports more timely HOs under dynamic channel conditions. In comparison, the fixed HOM and TTT parameters used by the baselines limit their ability to respond to dynamic channel conditions, potentially delaying necessary HOs to maintain connectivity. Nevertheless, RLFs are not entirely eliminated, since under substantial shadow fading considered in the urban scenario, unfavorable channel conditions could still cause link failures. 

Fig.~\ref{fig:cuHO} compares the cumulative HO delay of the three schemes. PPO-SHD incurs \SI{5.21}{\second}, only \SI{0.17}{\second} more than LD (\SI{5.04}{\second}) and 27.0\% less than O-HT (\SI{7.14}{\second}). Despite achieving the lowest HO delay, LD exhibits longer RLF duration and lower availability than PPO-SHD, indicating that minimizing HO delay alone does not ensure high service availability. PPO-SHD substantially reduces RLF duration at the cost of a modest increase in HO delay relative to LD. O-HT incurs higher HO delay because it favors the satellite with the highest RSRP without accounting for HO topology and its associated delay. These results demonstrate that PPO-SHD exploits topology information to balance HO delay and service continuity under dynamic channel conditions.

\subsection{Sensitivity to the Minimum
Elevation Angle}

% \begin{figure*}[!t]
%     \centering
%     \subfloat[]{%
%         \includegraphics[width=0.95\columnwidth]{figure/DiffAngle/availability_vs_ho_count.png}%
%         \label{ang_avail}%
%     }
%     \subfloat[]{%
%         \includegraphics[width=0.95\columnwidth]{figure/DiffAngle/availability_vs_rlf.png}%
%         \label{ang_bar}%
%     }
%     \caption{Performance of different angle. Availability with (a) HO count.
%     (b) RLF.}
%     \label{ang_fig}
% \end{figure*}
\begin{figure}[t]
    \centering
    \subfloat[\label{ang_avail}]{%
        \includegraphics[width=2.6in]{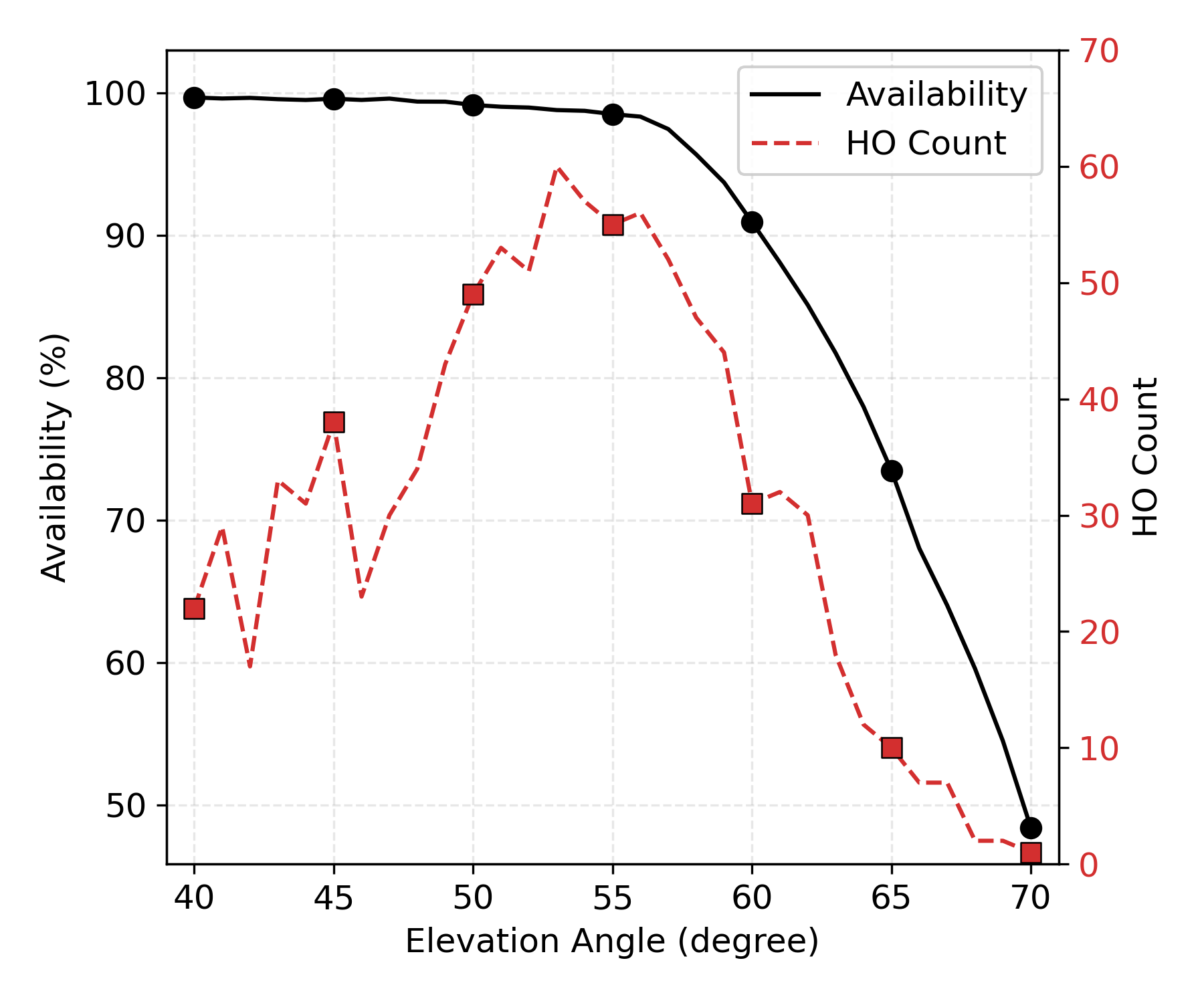}%
    }
    \par%\vspace{3mm}
       \vspace{-10pt}
    \subfloat[\label{ang_bar}]{%
        \includegraphics[width=2.6in]{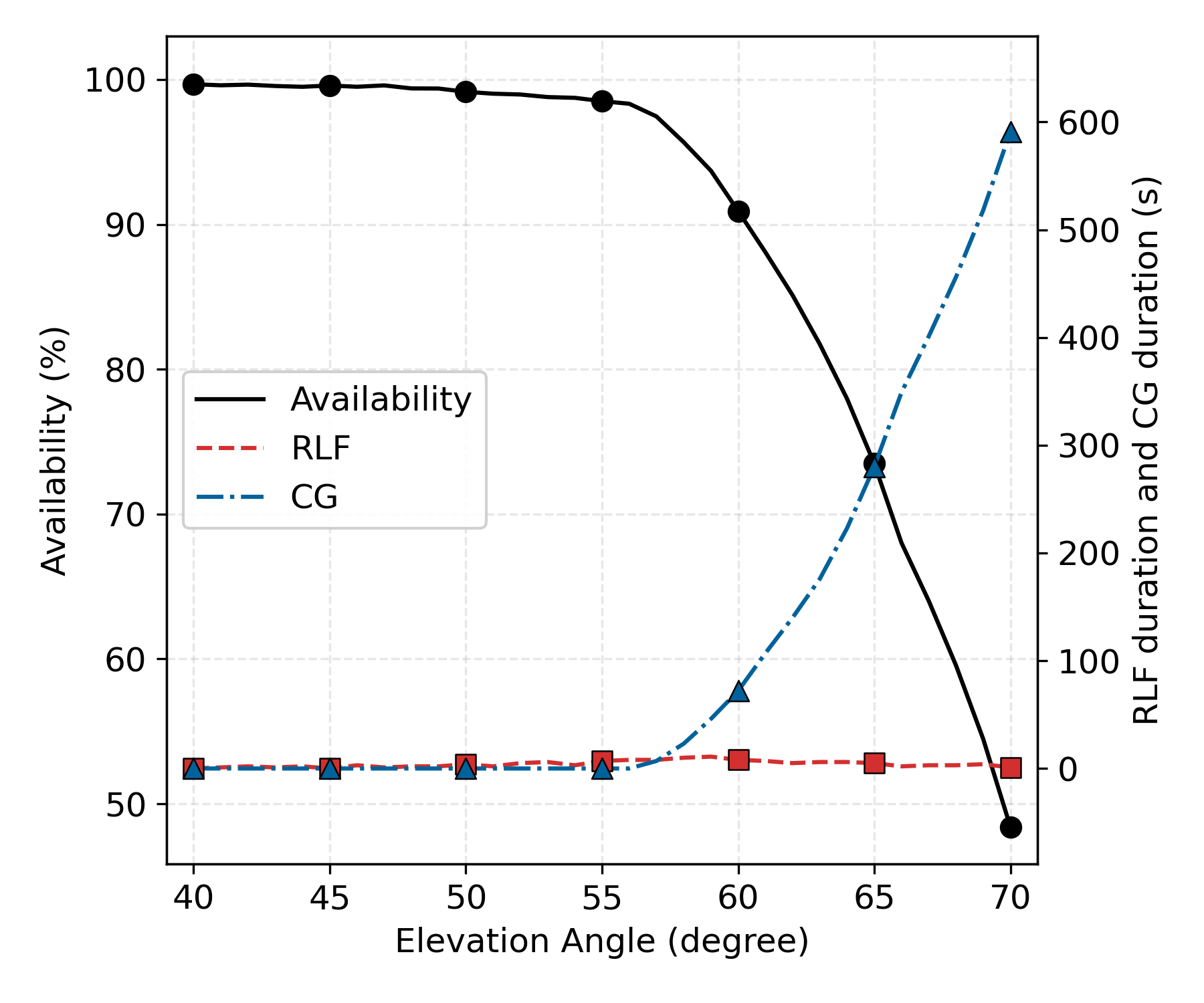}%
    }
    \caption{Effect of minimum elevation angle on availability with (a) HO counts, and (b) RLF duration and CG duration .}
    \label{ang_fig}
\end{figure}

Fig.~\ref{ang_fig} evaluates the PPO-SHD policy trained at $\theta_{\min}=40^\circ$ under minimum elevation angles ranging from $40^\circ$ to $70^\circ$, without retraining. In Fig.~\ref{ang_avail}, the HO count generally increases up to up to $55^\circ$, while availability decreases to 97.52\%. This trend is consistent with the shorter satellite visibility windows imposed by stricter elevation constraints, which increase the need for handovers to maintain connectivity. Beyond $55^\circ$, availability drops sharply to 47.44\% at $70^\circ$, despite a substantial reduction in the HO count.
Fig.~\ref{ang_bar} provides further insight into this degradation by comparing the availability associated with RLFs and CGs. The CG duration increases from zero at $55^\circ$ to \SI{591}{\second} at $70^\circ$, while the RLF duration remains comparatively small. The availability loss is therefore primarily associated with insufficient satellite coverage. Accordingly, the lower HO count in this regime reflects fewer opportunities for HO as satellite coverage becomes increasingly limited, rather than improved connection stability. Overall, PPO-SHD maintains high service availability under moderate increases in the minimum elevation angle within the evaluated deployment. However, its performance becomes increasingly constrained by candidate satellite availability at higher thresholds.

\begin{figure}[t]
    \centering
    \subfloat[\label{40deg}]{%
        \includegraphics[width=2.6in]{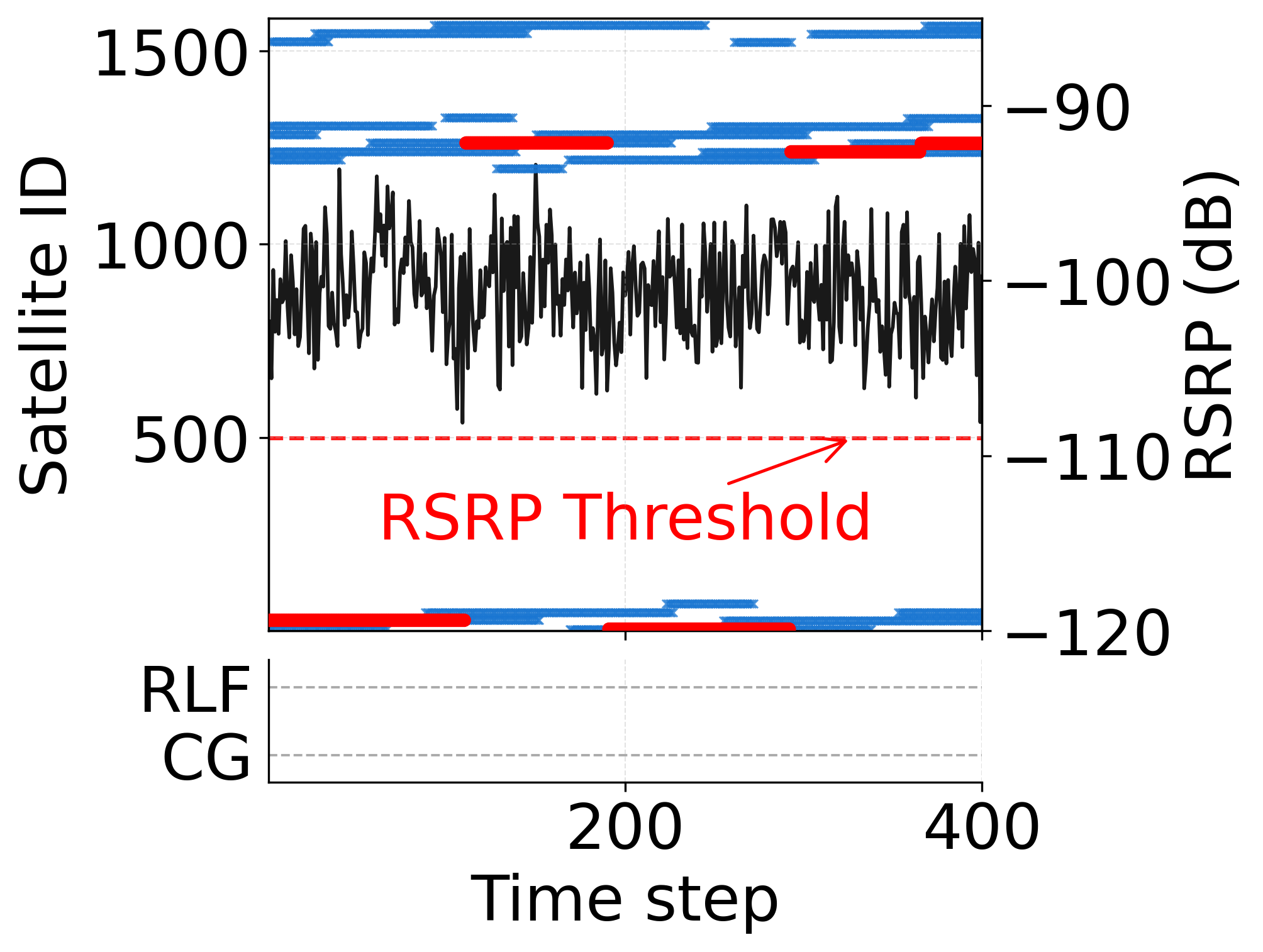}%
    }
    \par%\hspace{3mm}
       \vspace{-10pt}
    \subfloat[\label{65deg}]{%
        \includegraphics[width=2.6in]{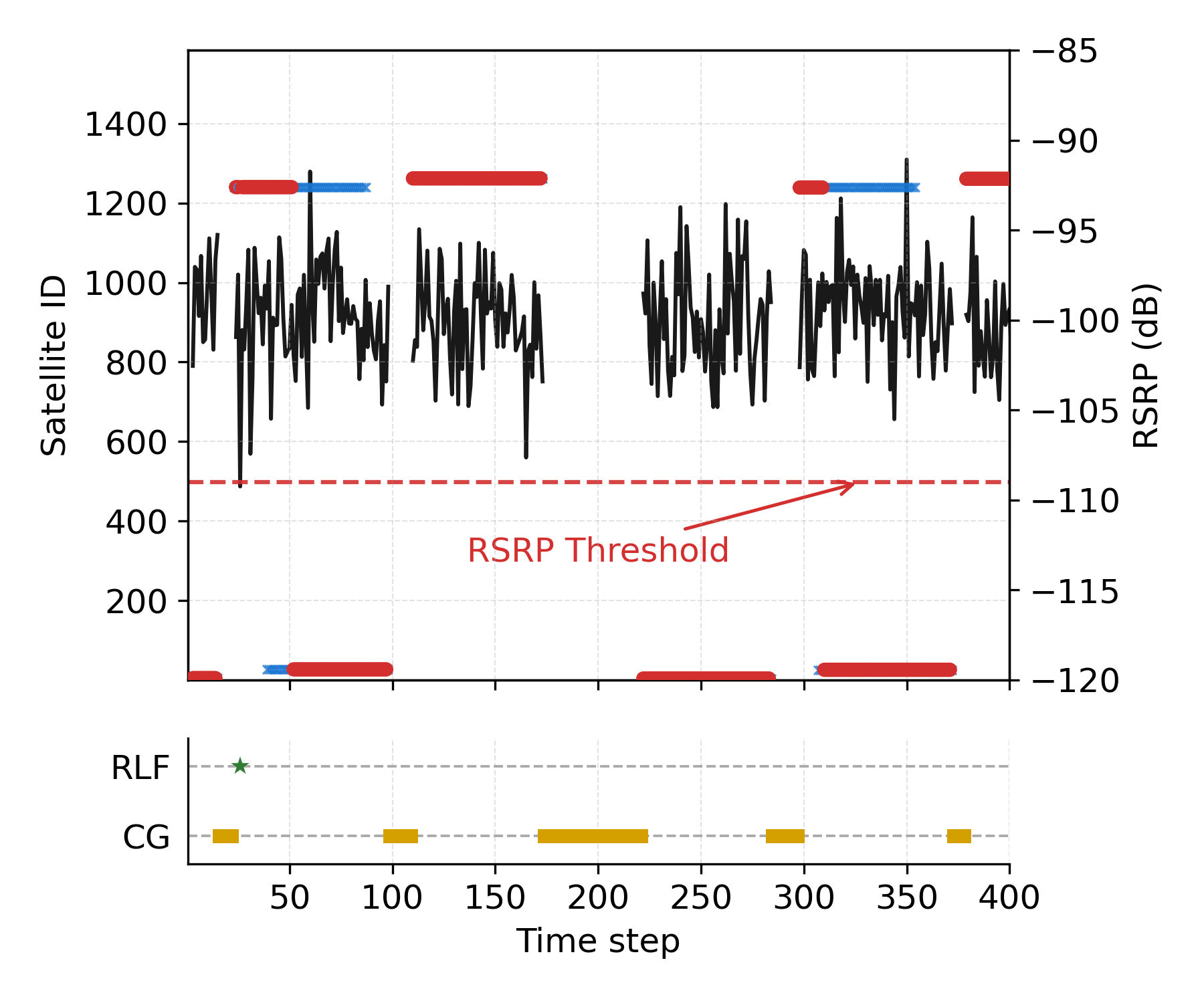}%
    }
    \caption{Satellite selection (red), candidate satellites (blue),  selected-link RSRP (black), and link continuity at $\theta_{min}$ of (a) $40^\circ$ and (b) $65^\circ$.}
    %Candidate satellites (blue), satellite selections , selected-link RSRP (black), and CG and RLF indicators over 400 slot evaluation intervals at minimum elevation angles  }
    \label{ang_sat}
\end{figure}

% \red{Fig.~\ref{ang_sat}  presents 400-slot excerpts from the 1200-slot evaluation trajectories at $\theta_{\min}=40^\circ$ and $65^\circ$, please improve the starting of this sentence these are veery good results but poorly described}
Fig.~\ref{ang_sat} illustrates the effectiveness of PPO-SHD under different satellite coverage conditions at $\theta_{\min}=40^\circ$ and $65^\circ$. As shown in Fig.~\ref{40deg}, PPO-SHD maintains connectivity throughout the 400-slot interval without RLFs while reducing HOs by retaining the serving satellite whenever feasible and triggering only necessary HOs. This behavior allows PPO-SHD to avoid unnecessary HO while preserving link continuity. In comparison, the more restrictive elevation requirement at $\theta_{\min}=65^\circ$ reduces the availability of eligible satellites and introduces intermittent CGs. Despite these constraints, PPO-SHD maintains connectivity during most periods when eligible candidates are available and re-establishes connections after the CGs. Although one RLF occurs during the observed interval, most interruptions coincide with CGs. These results highlight the ability of PPO-SHD to make timely and appropriate HO decision while preserving connectivity whenever satellite coverage permits.

\section{Conclusion}
\label{sec:con}
In this paper, we have developed a topology-aware PPO-based HO framework, PPO-SHD, to improve service continuity in gNB-onboard LEO satellite networks. The proposed framework incorporates RSRP, satellite–ground station associations, elevation angles, remaining visibility time, and association history to determine when to hand over and which satellite to select. In the evaluated urban scenario, PPO-SHD achieves a service availability of 99.32\% and reduces RLF duration by 85.7\% and 85.0\% relative to O-HT and LD, respectively. It also reduces cumulative HO delay by 27.0\% compared with O-HT while maintaining a delay comparable to LD. 
These results demonstrate the effectiveness of PPO-SHD in balancing HO delay and service continuity while highlighting the coverage limitations
imposed by candidate satellite availability. Future work could involve evaluating
its generalization across different constellation configurations and propagation environments.

% \section*{Acknowledgement}
% %\vspace{-1.5mm}

% This work was funded by the SSF research center, Sustainable Mobile Autonomous and Resilient 6G SatCom (SMART-6GSAT), under registry number CSG23-0001, and was also supported in part by the CELTIC-NEXT Project, 3D NETworks for 6G Mobile Communications Applications (3D-NET), with funding received from Vinnova, Swedish Innovation Agency with grant number 2025-02988.

\bibliographystyle{IEEEtran}
\bibliography{refs}

\end{document}